\documentclass[twocolumn,twoside,showpacs,prb]{revtex4}

\usepackage{graphicx}
   \graphicspath{{./img/}}
\usepackage{amsmath}
\usepackage{amssymb}
\usepackage{bm}
\usepackage{units}
\usepackage{wasysym}

\newcommand{\im}{i}	%\mathrm{i}}
\newcommand{\diff}{d}	%\mathrm{d}}
\newcommand{\e}{e}	%\mathrm{e}}
\newcommand{\version}{\today}
\DeclareMathOperator{\trace}{tr}

\begin{document}
%==========================================================================
\title{Incomplete pure dephasing of $N$-qubit entangled W states}
%==========================================================================

\author{Roland Doll}
\email{roland.doll@physik.uni-augsburg.de}
\author{Martijn Wubs}
\author{Peter H\"anggi}
\author{Sigmund Kohler}

\affiliation{%
Institut f{\"u}r Physik, Universit{\"a}t Augsburg, Universit\"atsstra{\ss}e
1,D-86135 Augsburg, Germany}
 
\date{\version}

\begin{abstract}
We consider qubits in a linear arrangement coupled to a bosonic field which acts
as a quantum heat bath and causes decoherence. By taking the spatial separation
of the qubits explicitly into account, the reduced qubit dynamics acquires an
additional non-Markovian element. We investigate the time evolution of an
entangled many-qubit W state, which for vanishing qubit separation remains
robust under pure dephasing. For finite separation, by contrast, the dynamics is
no longer decoherence-free. On the other hand, spatial noise correlations may
prevent a complete dephasing. While a standard Bloch-Redfield master
equation fails to describe this behavior even qualitatively, we propose instead
a widely applicable causal master equation. Here we employ it to identify and
characterize decoherence-poor subspaces. Consequences for quantum error 
correction are discussed.
\end{abstract}

\pacs{03.65.Yz, 63.22.+m, 03.67.Pp, 03.67.-a}

% ---------------------------------------------------------------------------

\maketitle

% ----------------------------------------------------------------------------
\section{Introduction}

In recent years, we witnessed great progress in the field of solid-state quantum
information processing, like the coherent control of single qubits
\cite{Nakamura1999a, Vion2002a, Chiorescu2003a} and two-qubit gates.\cite{Yamamoto2003a}
One of the major remaining challenges is decoherence: the
interaction of the qubits with their environment reduces the indispensable
quantum coherence and entanglement of the quantum states. This relates to the
scalability of the present few-qubit setups, because decoherence becomes more
pronounced as the number of qubits increases. Understanding the scaling of
multi-qubit decoherence is also experimentally relevant, as many groups take the
challenge of implementing more complex qubit architectures with solid-state
devices.

Not all many-qubit states are equally sensitive to the influence of an
environment. Depending on the symmetries of the qubits-environment coupling,
there can exist distinguished subspaces of a $N$-qubit Hilbert space that are
effectively decoupled from the environment and, thus, form so-called
decoherence-free subspaces (DFS).\cite{Palma1996a, Zanardi1997a, Lidar1998a,
Lidar2003a}
These allow one to implement logical decoherence-free qubits with two or several
physical qubits. Therefore, given an architecture for $N$ physical qubits, it is
essential to identify the sets of robust quantum states that suffer least from
decoherence.

\begin{figure}[t]
\includegraphics{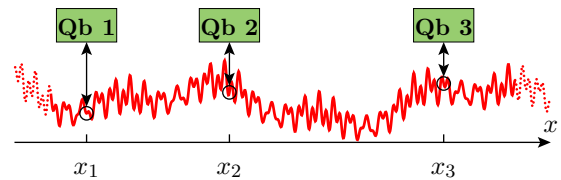}
\caption{\label{figCouplingToString}(Color online) Schematic representation of
$N$ qubits in a linear arrangement. The qubits (green boxes) 
interact via a coupling to the substrate phonon field (red line) 
at their positions $x_\nu$.}
\end{figure}

The substrate that supports the qubits also possesses its own degrees of freedom
like nuclear spins and phonons. They generally are coupled to the qubits and,
thereby, cause quantum dissipation and decoherence.\cite{Leggett1987a,
Hanggi1990a, Grifoni1998a, Gardiner1991a, Dittrich1998a, Breuer2002a} In experiments, one usually
observes both dephasing and relaxation, with dephasing happening on a faster
time scale.\cite{Borri2001a, Mozyrsky2002a, Muljarov2004a} It has been noted
that the reduced qubit dynamics in principle can be solved exactly if
the qubits experience pure phase noise,\cite{Luczka1990a,
vanKampen1995a, Unruh1995a, Palma1996a, Duan1998a, Krummheuer2002a,
Yu2002a, Yu2003a,Solenov2007a} and here we focus on this case as well. In
Ref.~\onlinecite{Doll2006a}, we recently presented
explicit expressions for the dephasing of two initially entangled qubits.  A
central conclusion of that work is that the entanglement of a robust entangled
state\nocite{Duer2000a,Eibl2004a,Roos2004a,Haeffner2005a}\cite{Yu2002a,robustentanglement}
is not perfectly stable but undergoes an initial decay,
stemming from the spatial qubit separation sketched in
Fig.~\ref{figCouplingToString}. This puts limitations on the applicability of
the concept of decoherence-free subspaces, since at best decoherence-poor
subspaces emerge instead.
%-------------
%-------------

Here we discuss the effects of phase noise on the $N$-qubit generalization of
the mentioned robust entangled state, namely the so-called W
state,\cite{Duer2000a} which is a coherent superposition of all states with
exactly one qubit in state $1$ while all the others are in state $0$ [cf.\
Eq.~\eqref{eqnRobustStateNQubits}]. The W states play an important role in
several protocols for quantum information processing, for example quantum
teleportation,\cite{Joo2003a,Agrawal2006a} superdense coding \cite{Agrawal2006a}
and quantum games.\cite{Han2002a} In quantum optics, the W states have already
been realized, first with three,\cite{Eibl2004a,Roos2004a,Mikami2005a} and
recently even with eight qubits.\cite{Haeffner2005a} Proposals exist to produce
W states in atomic gases \cite{Gorbachev2003a} and in solid-state
environments.\cite{Bruss2005a, Saito2006a, Wubs2007a}

In the limit of weak system-bath coupling, a successful and common approach to
quantum dissipation is provided by the Bloch-Redfield master
equation.\cite{Blum1996a}  A cornerstone of this formulation is neglecting
memory effects of the bath, so that one eventually obtains a Markovian master
equation.  The indirect interaction of two separated qubits via the environment,
however, introduces memory effects that arise when bath distortions can
propagate from one qubit to another during a finite time. This
timescale can be much larger than the intrinsic memory time of the bath. As will
be detailed below, a direct application of the Bloch-Redfield approach to
spatially separated qubits predicts spurious decoherence-free subspaces.

In the present work, we pursue a twofold goal: First, we consider an arbitrary
number of $N$ initially entangled qubits for which we present explicit
expressions for the coherence loss. This shows how in a linear arrangement
decoherence scales as a function of the number of qubits, and demonstrates the
consequences of spatial qubit separations. The stability of the W states
with respect to the system size $N$ has been studied previously,\cite{Carvalho2004a,Simon2002a,Duer2004a}
however for local decoherence models where the qubits couple to effectively
independent heat baths.
Second, by making a weak-coupling approximation, we derive a non-Markovian
master equation approach that captures the main effects of the spatial
separation. This approach has the advantage of being more intuitive, while still
allowing algebraic methods\cite{Lidar1998a,Lidar2003a} to be applied to the
problem of decoherence.  A comparison of the master-equation dynamics and the
exact dynamics therefore enables a better interpretation of the latter and a
critical examination of the validity of the former.  Moreover, the
master-equation approach will be applicable as well to other problems that do
not possess an exact solution.

The paper is organized as follows: In Sec.~\ref{secModel}, we present our model,
a system of $N$ qubits in a linear arrangement. The qubits interact with a
thermal bosonic field which causes decoherence of the $N$-qubit state. In
Sec.~\ref{secExact}, we present for pure dephasing the exact time evolution of
the concurrence for two qubits initially prepared in a robust entangled
Bell-state and then generalize our results to W states of an arbitrary number of
qubits.  Starting from the general model, we derive in Sec.~\ref{secMasterEqn} a
master equation for the reduced dynamics, taking special care of the spatial
qubit separation. Its solution is then compared with the exact results for pure
dephasing. The derivations of the exact solution and of a
convolutionless master equation are deferred to two appendices.

%------------------------------------------------------------------------------
\section{Qubits coupled to a bosonic field}\label{secModel}

As a model for the $N$ qubits in the bosonic environment sketched in
Fig.~\ref{figCouplingToString}, we employ the Hamiltonian
\begin{equation}
\label{eqnHamiltonian}
H = H_0 + H_\mathrm{qb}\,,
\end{equation}
where the qubits and the bosonic field in the absence of the coupling
are described by the Hamiltonian\cite{Leggett1987a, Hanggi1990a,
Grifoni1998a}
\begin{equation}
\label{eqnH0}
H_0 = \frac{\hbar}{2} \sum_{\nu = 1}^N \Omega_\nu \sigma_{\nu z}
   + \sum_k \hbar \omega_k b_k^\dagger b_k\,.
\end{equation}
The first term in Eq.~\eqref{eqnH0} represents $N$ qubits $\nu =
1,2,\ldots,N$ with energy splittings $\hbar \Omega_\nu$ and Pauli
matrices $\sigma_{\nu z}$.
Since we will not address the coherent control of individual qubits
explicitly, the specific choice for the energy splittings is not of
major relevance. Note that there is no direct interaction between the
qubits.
The second term in Eq.~\eqref{eqnH0} describes a bosonic field that
consists of modes $k$ with energies $\hbar\omega_k$ and the usual
bosonic annihilation and creation operators $b_k$ and $b_k^\dagger$.
We restrict ourselves to a linear dispersion relation $\omega_k= c|k|$
with $c$ being the sound velocity.

Qubit $\nu$ is located at position $x_\nu$ and couples linearly via the
operator $X_\nu$ to the field, so that the coupling Hamiltonian reads
\begin{equation} \label{eqnCoupling}
H_\mathrm{qb} =  \hbar \sum_{\nu = 1}^N X_\nu \xi_\nu\,,
\end{equation}
with
\begin{equation}
\label{eqnFieldOperator}
\xi_\nu = \xi(x_\nu) = \sum_k g_k \e^{\im k x_\nu} (b_k +b_{-k}^\dagger)
\end{equation}
the bosonic field operator at position $x_\nu$. We assume the coupling strengths
to be real, isotropic, and identical for all qubits, i.e.\ $g_{k\nu} = g_k$ and
$g_{-k} = g_k$.

We choose an initial condition of the Feynman-Vernon type, i.e.~at time $t=t_0$,
the bath is at equilibrium and is not correlated with the qubits. Thus the total
initial density matrix $R(t_0)$ is a direct product of a qubit and bath density
operator,
\begin{equation}\label{eqnInitialCondition}
R(t_0) = \rho(t_0)\otimes\rho^\mathrm{eq}_\mathrm{b}\,.
\end{equation}
Here, $\rho$ is the reduced density matrix of the qubits and
\begin{equation}\label{rhobath}
\rho^\mathrm{eq}_\mathrm{b} = \frac{1}{Z} \exp\bigg(-\sum_k \frac{\hbar\omega_k
b_k^\dagger b_k}{k_\mathrm{B}T}\bigg)
\end{equation}
is the canonical ensemble of the bosons at temperature $T$ and $Z$ is
the corresponding partition function.

The dynamics of the qubits plus the environment is governed by the Liouville-von
Neumann equation
\begin{equation} \label{eqnLiouvillevonNeumann}
\im\hbar\frac{\diff}{\diff t} \widetilde R(t)
= \bigl[\widetilde H_\text{qb}(t), \widetilde R(t)\bigr]\,.
\end{equation}
The tilde denotes the interaction-picture representation with respect to $H_0$,
i.e.\ $\tilde A(t) = U_0^\dagger(t) A\, U_0(t)$, where $U_0(t) = \exp\{-i
H_0(t-t_0)/\hbar\}$.
We are exclusively interested in the state of the qubits, so our goal is to find
the time evolution of the reduced density operator
$\tilde\rho(t) = \trace_\mathrm{b} \tilde R(t)$, where $\trace_\mathrm{b}$
denotes the trace over the environmental degrees
of freedom.  In what follows, we will consider the density matrix elements
$\tilde\rho_{\mathbf{m},\mathbf{n}} =
\langle\mathbf{m}|\tilde\rho|\mathbf{n}\rangle$ in the basis $|\mathbf{n}\rangle
= |n_1,n_2,\ldots,n_N \rangle$, where $\sigma_{\nu z} |\mathbf{n}\rangle =
(-1)^{n_\nu} |\mathbf{n} \rangle$ and $n_\nu = 0,1$.

%------------------------------------------------------------------------------
\section{Exact reduced dynamics}\label{secExact}

Generally, the coupling of a qubit to an environment induces spin flips and also
randomizes the relative phase between the eigenstates of the qubit. If the
coupling operator $H_\text{qb}$ commutes with the qubit Hamiltonian, the qubits
experience the so-called pure phase noise. Consequently, one finds
$[H_\text{qb},U_0]=0$ so that the interaction-picture qubit operators remain
time-independent, $\widetilde X_\nu(t) = X_\nu$. In particular, the
time-evolution of the reduced density operator $\tilde\rho$ is independent of
the coherent oscillation frequencies $\Omega_\nu$ of the qubits.

In the following, we consider the coupling operators $X_\nu =\sigma_{\nu z}$
which constitute a case of pure phase noise. The reduced qubit dynamics can then
be solved analytically. We defer the explicit derivation to
Appendix~\ref{secDerivationExactSolution}, where we obtain
\begin{equation} \label{eqnExactSolution}
\tilde\rho_{\mathbf{m},\mathbf{n}}(t) = \rho_{\mathbf{m},\mathbf{n}}(0)
\,e^{-\Lambda_{\mathbf{m},\mathbf{n}}(t) + \im
\phi_{\mathbf{m},\mathbf{n}}(t)} ,
\end{equation}
with the amplitude damping\nocite{Doll2006a}\cite{misprintEPL}
\begin{equation} \label{eqnDamping}
\begin{split}
\Lambda_{\mathbf{m},\mathbf{n}}(t)
&= \int_0^\infty\!\diff\omega\, J(\omega)
  \frac{1-\cos(\omega t)}{\omega^2} \coth\left(
  \frac{\hbar \omega}{2k_\text{B}T}\right) \\
&\quad \times
  \bigg|\sum_{\nu=1}^N [(-1)^{m_\nu} - (-1)^{n_\nu}] \e^{\im \omega t_\nu}
  \bigg|^2  ,
\end{split}
\end{equation}
and the time-dependent phase shift
\begin{equation} \label{eqnPhaseShift}
\begin{split}
\phi_{\mathbf{m},\mathbf{n}}(t) &= \int_0^\infty \! \diff\omega\, J(\omega)
\frac{\omega t-\sin(\omega t)}{\omega^2}\\ 
&\quad \times  \sum_{\nu,\nu^\prime = 1}^N  [(-1)^{m_\nu+m_{\nu^\prime}} -
(-1)^{n_\nu+n_{\nu^\prime}}]\cos(\omega t_{\nu\nu^\prime}).
\end{split}
\end{equation}
For ease of notation, we have set the initial time $t_0 = 0$ and introduced the bath
spectral density $J(\omega) = \sum_k g_k^2 \delta(\omega - ck)$. The transit
time of a wave between the qubits $\nu$ and $\nu'$ reads $t_{\nu\nu'} =
x_{\nu\nu'}/c$, where $x_{\nu\nu'} = |x_\nu-x_{\nu'}|$. We will focus on a
linear arrangement of the $N$ qubits and consider equal
nearest-neighbor separations $x_{\nu,\nu+1}=x_{12}$. Then the transit time
becomes $t_{\nu\nu^\prime} = |\nu-\nu^\prime|t_{12}$. 

To elaborate on the impact of spatially correlated noise, we assume the chain of
qubits to be embedded in a medium with a channel structure, i.e.~we treat 
the bosonic field as effectively one-dimensional. Quasi one-dimensional
geometries may, for example, be realized by carbon nanotubes or linear ion
traps. They represent configurations in which the requirements of qubit protection and 
adressability are well-balanced, and where we expect the effects of spatial 
noise correlations to be most overt. In this case, the spectral density is of 
the ohmic type,\cite{Leggett1987a}
\begin{equation}\label{ohmic}
J(\omega) = \alpha\, \omega e^{-\omega/\omega_\text{c}} ,
\end{equation}
wherein $\omega_\text{c}$ denotes a cutoff frequency which for a phonon field is
the Debye frequency. For a more detailed discussion on the relation of the 
spectral density to the substrate geometry, in particular in the context of
semiconductor quantum dots, we refer the reader to
Refs.~\onlinecite{Fedichkin2006a,Solenov2007a}.

For a physical realization with a GaAs substrate, the Debye frequency is of the
order $\omega_\text{c}=\unit[5\times 10^{13}]{Hz}$ while the sound velocity is
$c=\unit[3\times10^3]{ms^{-1}}$. Then a qubit separation $x_{12}=
\unit[100]{nm}$ corresponds to the transit time $t_{12} = 10^4/\omega_\text{c}$.
Likewise, for a temperature $T = \unit[10]{mK}$ we have $k_\text{B} T / \hbar
\omega_\text{c} = 10^{-4}$. This implies that $\omega_\text{c}^{-1}$ is
typically the smallest timescale of the problem, while the transit time and the
thermal coherence time can be of the same order. The present work is devoted to
the effects of spatial separation between qubits, but not of the finite extent
of the qubits themselves. Note that in semiconductor quantum dots, the finite
width $a$ of an electron wavefunction in the dot (typically below 100
nanometers) may lead to an effective cutoff frequency $\omega_\text{c} \approx
c/a$ that is smaller than the Debye frequency, but still larger than all other
frequencies in the system.\cite{Brandes2002a,Fedichkin2004a,Vorojtsov2005a}

Both the damping~\eqref{eqnDamping} and the phases~\eqref{eqnPhaseShift} vanish
at time $t=0$, so that Eq.~\eqref{eqnExactSolution} is consistent with the
initial condition. As expected for pure dephasing, populations are preserved,
i.e.\ the diagonal matrix elements obey $\tilde\rho_\mathbf{m,m}(t) =
\tilde\rho_\mathbf{m,m}(0)$.  This implies that generally neither the qubits nor
the total system will reach thermal equilibrium.  However, the relative phases
between eigenstates will be randomized so that off-diagonal density matrix
elements --- the so-called coherences --- may decay, which reflects the
process of decoherence.

\subsection{Dephasing of robust entangled states}
A most relevant decoherence effect in a quantum computer is the loss
of entanglement between different qubits.  In order to exemplify the
impact of a spatial qubit separation on decoherence, we consider as
the initial state the robust entangled $N$-qubit W state
  \begin{equation}\label{eqnRobustStateNQubits}
  |W_N\rangle = 
  \frac{1}{\sqrt{N}}\bigl(\, |100 \ldots 0\rangle
   + |010 \ldots 0\rangle + \ldots
   + |000 \ldots 1\rangle\,\bigr).
  \end{equation}
For two qubits ($N=2$) it has been
shown that the two-qubit entanglement inherent in the state $|W_2\rangle$, which
is the symmetric Bell state, is robust under dephasing for vanishing spatial
separation,\cite{Yu2002a,Yu2003a} while it decays for finite
separation.
\cite{Palma1996a, Doll2006a}

Our motivation to focus on the initial
states \eqref{eqnRobustStateNQubits} is twofold: First, W states play an
important role in several protocols for quantum information processing,\cite{Joo2003a,Han2002a,Agrawal2006a}
so that their sensitivity to an environment is relevant in itself. Second,
among all fully entangled $N$-qubit states the $W$ states are special
in that they maintain their $N$-qubit entanglement under collective
dephasing (i.e.\ for vanishing qubit separations). Now if already the
W states start to loose their entanglement due to a finite spatial
separation of the qubits, then this is a strong indication
that for other fully entangled $N$-qubit states, the situation would
be worse.  Or, to put it simply, we wish to give the most
optimistic estimate about $N$-qubit decoherence and to our knowledge
the best way to do that is by focusing on the W states.  

The fact that no bit flips occur under pure dephasing is reflected in
the structure of the exact solution \eqref{eqnExactSolution}: 
All density matrix elements that are initially zero remain zero, so
that for the state $|W_N\rangle$, the dissipative quantum dynamics is
restricted to the states
\begin{equation} \label{eqnMatrixElementOneQubitExited}
|j\rangle = |0 0\ldots 1_{j} \ldots 0\rangle, \quad j = 1,2,\ldots,N.
\end{equation}
Thus at most $N^2$ out of $2^{2N}$ density matrix elements are
nonvanishing. Initially they are all equal,
i.e.~$\rho_{jj^\prime}(0)=1/N$.
From the Hamiltonian~\eqref{eqnH0} it directly follows, that the
states $|j\rangle$ possess the eigenenergies $\hbar\omega_{j} =
\frac{\hbar}{2}\sum_{\nu=1}^{N}\Omega_{\nu} - \hbar\Omega_{j}$ 
and a back-transformation of the coherence $\tilde \rho_{jj'}(t)$ 
to the Schr\"odinger picture provides the phase factor 
$\exp[i(\omega_j-\omega_{j'})t]$.

\paragraph{Frequency shifts.}
One effect of the coupling to a heat bath is a frequency shift
$\delta\Omega_j$ which we obtain in the following way: Upon noticing that one
can separate the phases \eqref{eqnPhaseShift} into terms that depend on only $j$
or $j^\prime$, we write $\phi_{jj^\prime}(t) = \phi_{j}(t) -
\phi_{j^\prime}(t)$. Each $\phi_j(t)$ turns out to consist of a finite
contribution and a contribution that grows linearly in time,
i.e.~$\phi_j(t) = \varphi_j(t)-\delta\Omega_j\,t$. The latter leads
to a (static) frequency shift.
For the ohmic spectral density~\eqref{ohmic}, we obtain
\begin{align}\label{eqnFrequencyShiftStatic}
\delta\Omega_{j}
&= - \alpha \sum_{\nu,\nu'=1}^{N} (-1)^{\delta_{j\nu}+\delta_{j\nu'}}
   \frac{\omega_\text{c}}{1 + \omega^2_{c}t^2_{\nu\nu'}}\,,
\\ \label{eqnFrequencyShiftDynamic}
\varphi_j(t)
&= -\frac{\alpha}{2}
   \sum_{\nu,\nu'=1}^{N} (-1)^{\delta_{j\nu}+\delta_{j\nu'}}
   \sum_{\pm}\arctan[\omega_\text{c} (t\pm t_{\nu\nu'})] \,.
\end{align}
Thus the effective energy splitting of qubit $j$ becomes $\hbar(\Omega_{j} +
\delta\Omega_{j})$.  Note that both $\delta\Omega_j$ and $\varphi_j(t)$ depend
on the transit times $t_{\nu\nu'}$ and the system size $N$ but not on the
temperature. They can be interpreted as a result from an effective coherent
interaction of the qubits mediated by the vacuum fluctuations of the bosonic
field,\cite{Solenov2007a} where $\delta\Omega_j$ arises from an induced
static exchange interaction and its onset is described by $\varphi_j(t)$.
Note that the dominant contribution to the static shift
$\delta\Omega_j$ stems from the diagonal terms $\nu=\nu'$ in
Eq.~\eqref{eqnFrequencyShiftStatic}, whereas the non-diagonal terms are
suppressed by a factor $\omega_\text{c}^2 t_{\nu\nu'}^2$, respectively.

We henceforth work in the interaction picture with respect to the renormalized
energies so that the density matrix element $\tilde\rho_{jj'}$ reads
\begin{equation} \label{eqnMatrixElementsRobustStateRenormalized}
\tilde\rho_{jj'}(t) = \frac{1}{N}e^{-\Lambda_{jj'}(t)+i\varphi_{jj'}(t)}.
\end{equation}
The time-dependent phases $\varphi_{jj'}(t) = \varphi_j(t) - \varphi_{j'}(t)$
decay to zero after a rather short time $t \simeq \omega_\text{c}^{-1}$ and, thus,
influence the decoherence process only during a short initial stage.

\paragraph{Damping factors.}
The coherence loss is given by the damping factors $\exp[-\Lambda_{jj'}(t)]$.
Inserting the ohmic spectral density~\eqref{ohmic} into expression
\eqref{eqnDamping}, we obtain for them the explicit form
\begin{widetext}
\begin{equation}\label{exactexplicit}
\e^{-\Lambda_{jj^\prime}(t)} = \frac{1}{N} \left| \frac{\Gamma\bigl(\frac{k_\text{B}T}{\hbar \omega_\text{c}} [1 - \im
\omega_\text{c} t_{jj'}]\bigr)}{\Gamma\bigl(\frac{k_\text{B}T}{\hbar
\omega_\text{c}}\bigr)} \right|^{16 \alpha}
\hspace{-0ex}\times\hspace{-0.0ex}
\left| \frac{\Gamma^2\bigl( \frac{k_\text{B}T}{\hbar \omega_\text{c}}[1 + \im
\omega_\text{c} t]\bigr)\,\Gamma^2\bigl( \frac{k_\text{B}T}{\hbar
\omega_\text{c}}[1 -\im \omega_\text{c}t ]) \, (1 + \omega_\text{c}^2 t^{2}) }{
\Gamma^2\bigl(\frac{k_\text{B}T}{\hbar \omega_\text{c}} [1 + \im\omega_\text{c}
(t-t_{jj^\prime})]\bigr)\, \Gamma^2\bigl( \frac{k_\text{B}T}{\hbar
\omega_\text{c}}[1 - \im \omega_\text{c} (t+t_{jj^\prime})]\bigr) \, \left( 1 +
\frac{\omega_\text{c}^2 t^{2}}{(1 - \im \omega_\text{c} t_{jj^\prime})^2}
\right) }  \right|^{4 \alpha}\hspace{-2ex},
\end{equation}
\end{widetext}
where $\Gamma$ denotes the Euler Gamma function. The nominator of the second
factor in Eq.~\eqref{exactexplicit} itself is already of physical interest: It
describes the decoherence of a single qubit in the absence of the other
qubits.\cite{Doll2006a} One can identify three stages in the single-qubit time
evolution:\cite{Palma1996a} Very shortly after the preparation, i.e.\ for times
$t \apprle \omega_\text{c}^{-1}$, the fluctuations of the bosonic field are not
yet effective, leading to a ``quiet regime'' in which essentially no
single-qubit decoherence takes place. At an intermediate stage,
$\omega_\text{c}^{-1} \apprle t \apprle \hbar / k_\text{B} T$, the main origin
of single-qubit decoherence is  vacuum quantum fluctuations. They lead to an
initial slip of the coherence which we discuss in detail below. Finally for
times larger than the thermal coherence time, $t \apprge \hbar/k_\text{B} T$,
thermal fluctuations dominate the coherence loss. The dephasing will finally be
complete, i.e.\ a single qubit that starts in a superposition will loose all
quantum coherence due to dephasing caused by the one-dimensional
bath.\cite{Doll2006a}

For the spatially separated qubits prepared in the W state
\eqref{eqnRobustStateNQubits} that we focus on here, the transit times
$t_{jj'}$ introduce additional time scales after which the denominator of the
second factor in Eq.~\eqref{exactexplicit} becomes relevant. Since the
denominator ultimately decays equally fast as the nominator, decoherence
will come to a standstill. This interesting behavior does not occur for a single
qubit and will now first be investigated for the case of a qubit pair.

\subsection{Entanglement of qubit pairs}\label{secTwoQubits}
It is instructive to first consider the case of two qubits, $N=2$,
which represents the most basic system in which a spatial qubit
separation influences quantum coherence. The robust initial state
\eqref{eqnRobustStateNQubits} then reduces to the maximally entangled
Bell state
\begin{equation} \label{eqnRobustState2Qubits}
|W_2\rangle = \frac{1}{\sqrt{2}}\bigl(|10\rangle +
|01\rangle\bigr)=\frac{1}{\sqrt{2}}\bigl(|1\rangle + |2\rangle\bigr)\,,
\end{equation}
where the last expression refers to the basis
\eqref{eqnMatrixElementOneQubitExited}.

For a bipartite system, the degree of entanglement can be measured by
the concurrence\cite{Wootters1998a}
\begin{equation}
C
\equiv \max\Bigl\{0,
\sqrt{\lambda_1}-\sqrt{\lambda_2}-\sqrt{\lambda_3}-\sqrt{\lambda_4}\Bigr\},
\end{equation}
where the $\lambda_i$ denote the eigenvalues of the matrix
$\rho\sigma_{y1}\sigma_{y2}\rho^\ast\sigma_{y1}\sigma_{y2}$ in
decreasing order and $\rho^\ast$ is the complex conjugate of $\rho$.  For
qubit pairs that are initially prepared in state
\eqref{eqnRobustState2Qubits} and subject to pure phase noise, the
concurrence can be expressed at all times by the absolute value of a
single non-diagonal element, namely $C(t) = 2|\rho_{01,10}(t)|
=2|\rho_{12}(t)|$, irrespective of the spatial separation.\cite{Doll2006a}

\begin{figure}
\includegraphics{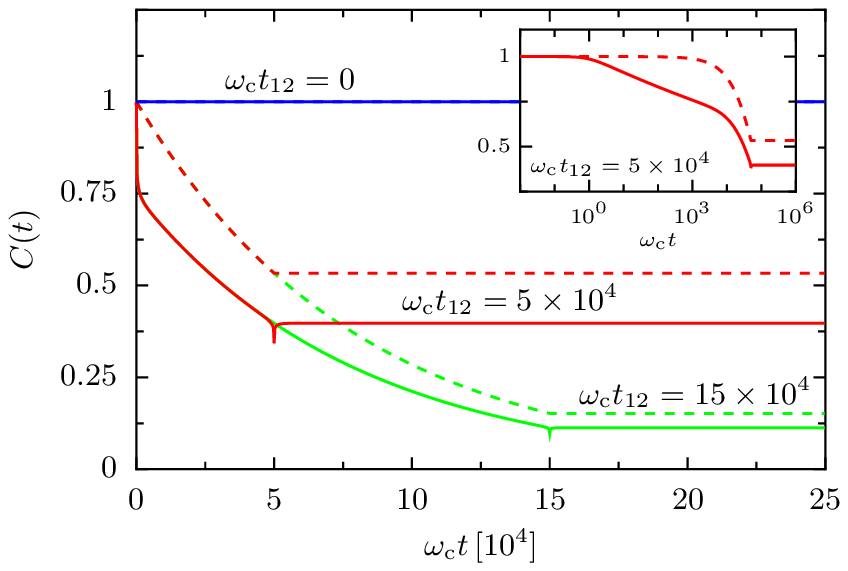}
\caption{\label{figCExactPhaseNoise}(Color online) Time evolution of the
concurrence $C$ for two qubits initially prepared in the robust
state~\eqref{eqnRobustState2Qubits} for various transit times
$t_{12}$: exact time evolution (solid lines) compared to the results
obtained from the causal master equation derived in
Sec.~\ref{secMasterEqn} (dashed).
The temperature is $k_\mathrm{B}T = 10^{-4}\hbar\omega_\mathrm{c}$ and
the coupling strength $\alpha = 0.005$.  For $\Omega =
10^{-3}\omega_\mathrm{c} $, the time range corresponds to 40
coherent oscillations.
Inset: Blow-up on a logarithmic scale for the transit time
$t_{12} = 5\times 10^4/\omega_\text{c}$.}
\end{figure}

Figure~\ref{figCExactPhaseNoise} shows the time dependence of the
concurrence of a qubit pair prepared in the robust entangled state
\eqref{eqnRobustState2Qubits} for various spatial separations.
For vanishing separation, we find the concurrence $C(t)=1$, which means
that the entanglement indeed remains perfect and justifies
the designation ``robust entangled state''.\cite{Yu2002a, Yu2003a}
This is tantamount to saying that the state $|W_2\rangle$ is an
element of a so-called decoherence-free subspace of the two-qubit
Hilbert space. However, Fig.~\ref{figCExactPhaseNoise} also shows that
the concurrence of the robust state {\em does} decay if the qubit
separation $x_{12}$ is finite. The decay lasts until the transit time $t_{12}$
is reached. From then on, the concurrence remains constant, so that both the
entanglement and the coherence become stable.
Thus, we can conclude that for spatially separated qubits the state $|W_2\rangle$ is not an element of a
decoherence-free subspace, but rather of a decoherence-poor subspace.
Its emergence from the exact solution \eqref{exactexplicit} will be
discussed for the more general case of $N$ qubits in the subsequent section.
Moreover, a more intuitive picture will be drawn in the framework of a
causal master equation approach in Sec.~\ref{secMasterEqn}.
%-------------
%-------------

\begin{figure}
\includegraphics{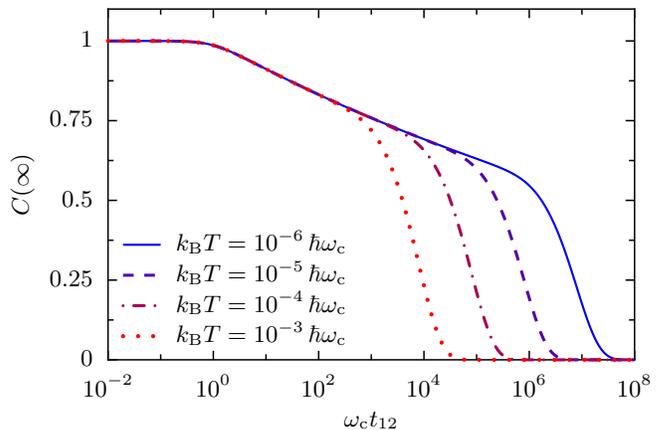}
\caption{\label{figCssExactPhaseNoise}(Color online) Final value of
concurrence for the robust state $|W_2\rangle$ as a function of the
spatial separation $x_{12}= c\,t_{12}$ for various temperatures.
The coupling strength is $\alpha = 0.005$.}
\end{figure}

Further information is provided by the value to which the concurrence
saturates.  Figure~\ref{figCExactPhaseNoise} shows that
$C(t\to\infty)$ is influenced by both the transit time between the
qubits and the properties of the bosonic field. The exact value can be
obtained from Eq.~\eqref{exactexplicit} and reads
\begin{equation} \label{eqnSaturatedConcurrence}
C(\infty) = (1 + \omega_\mathrm{c}^2t_{12}^2)^{4\alpha} \left|
\frac{\Gamma\left(
 \frac{k_\mathrm{B} T}{\hbar\omega_\mathrm{c}}[1-\im \omega_\mathrm{c} t_{12}]
 \right)}{\Gamma\left(\frac{k_\mathrm{B} T}{\hbar\omega_\mathrm{c}}\right)}
\right|^{16\alpha} .
\end{equation}
Figure~\ref{figCssExactPhaseNoise} shows this final concurrence $C(\infty)$ as a function
of the transit time $t_{12}$ for various temperatures $T$. 
As for the time-evolution, three regimes can be identified: For the
(unphysically small) separations $x_{12}<c/\omega_\text{c}$, the
concurrence remains at $C=1$, while for $c/\omega_\text{c}< x_{12} < \hbar
c/k_\text{B}T$, the entanglement is no longer perfect, but still at an
appreciably large value. For large separations, $x> \hbar c/k_\text{B}T$,
the concurrence essentially decays to zero. The latter limit is a
prerequisite for the application of quantum error-correction schemes that
assume that the qubits experience uncorrelated noise.
%-------------
%-------------

Summarizing our two-qubit results, we find that the two-qubit concurrence can
saturate at a stable value as long as the separation of the qubits stays finite,
i.e.~as long as $x_{12} < \infty$. Remarkably, the coherence of a single qubit
in the same environment (i.e.~when only one qubit is present) decays completely.
The entanglement dynamics that we find for the robust-entangled Bell state
\eqref{eqnRobustState2Qubits} is highly non-Markovian and it is obvious that
such a behavior can not be described by a Markovian master equation such as the
standard Bloch-Redfield equations, as we discuss below. This fact, and in particular the final saturation of
concurrence as shown in Figs.~\ref{figCExactPhaseNoise} and
\ref{figCssExactPhaseNoise} contradict expectations formulated in
Ref.~\onlinecite{Solenov2007a}; calculations in that work involve
non-entangled initial states, while we study robust-entangled Bell states.

Finally, we want to note that an entanglement saturation under pure dephasing
may appear for baths with super-ohmic spectral densities as
well,\cite{Krummheuer2002a,Roszak2006a,Doll2006a} even in the limit of large
qubit separations. This behavior is in contrast to the intuition emerging 
from recent work,\cite{Yu2002a,Yu2003a} namely that the entanglement
would always decay to zero when qubits are coupled to independent heat baths.
For super-ohmic baths in higher dimensions, a finite final entanglement
does not necessarily stem from spatial correlations. This becomes obvious
from the fact that for super-ohmic baths as for example a three-dimensional 
phonon field, the  coherence of {\em single} qubits exhibits a similar saturation.

\subsection{$N$-qubit fidelity}\label{secNQubits}
Let us now turn to the intriguing question how the previous results
can be generalized to arrays of qubits. As already described above,
the focus will be on the scaling of the decoherence as a function of
the system size $N$ of linearly arranged qubits. In particular, we
study how the amount of entanglement evolves for the qubits that start
in the $N$-qubit W state~\eqref{eqnRobustStateNQubits} and, due to
dephasing, at later times must be described by the $N$-qubit density
matrix \eqref{eqnExactSolution}.

With the exact dynamics known, the only remaining question is how to quantify
the entanglement. If there were no interaction with the environment, then the
qubits would remain in their pure entangled  W
state~(\ref{eqnRobustStateNQubits}), which in density-matrix notation reads
$\rho(0) = |{W_{N}}\rangle\langle{W_{N}}|$. After the dissipative
time evolution, the qubit state deviates from this ``ideal'' output state
$\tilde\rho_\text{ideal}(t) = \rho(0)$. The question is how much. A proper
measure for this quantity is the fidelity\cite{Poyatos1997a} $F(t) =\trace\{
\rho(t) \rho_\text{ideal}(t)\}$, which in our case reads
\begin{equation} \label{eqnFidelity}
F(t) = \trace \left\{ \rho(0)\tilde\rho(t) \right\}
= \langle W_N|\tilde\rho(t)|W_N\rangle.
\end{equation}
In general, the fidelity is bounded by $0 \le F \le 1$, where $F=1$ corresponds
to a pure state.\cite{Nielsen2000a} For qubits subject to pure dephasing, the
somewhat more strict condition $\sum_{j}\rho_{jj}^{2}(0) \le F \le 1$
applies, because the populations do not change. In particular the inequalities
$1/N \le F \le 1$ will hold for the initial state $|{W_{N}}\rangle$, as
illustrated below. To give another argument why fidelity makes a good measure of
entanglement for our particular purpose, we emphasize that for two
qubits prepared in the state \eqref{eqnRobustState2Qubits}, the
fidelity directly relates to the concurrence via the relation $C(t) =
2F(t)-1$.

We note in passing that for other initial states that are {\em not} pure
$N$-qubit entangled states, one should be careful to use fidelity as an
entanglement measure, for example because the fidelity may remain constant while
the system undergoes nontrivial dynamics.\cite{Fedichkin2006a} Finding other
entanglement measures for three or more qubits is an active field of
research.\cite{Duer2000a,Mintert2005a,Lohmayer2006a,Horodecki2007a} Their numerical evaluation can be
rather involved, especially for larger systems. These issues need not concern us
here, since we start with an $N$-qubit entangled pure state for which the
fidelity is a good measure of entanglement. The fidelity has the additional
advantage that it is easily evaluated analytically for larger systems as well. 

\begin{figure}
\includegraphics{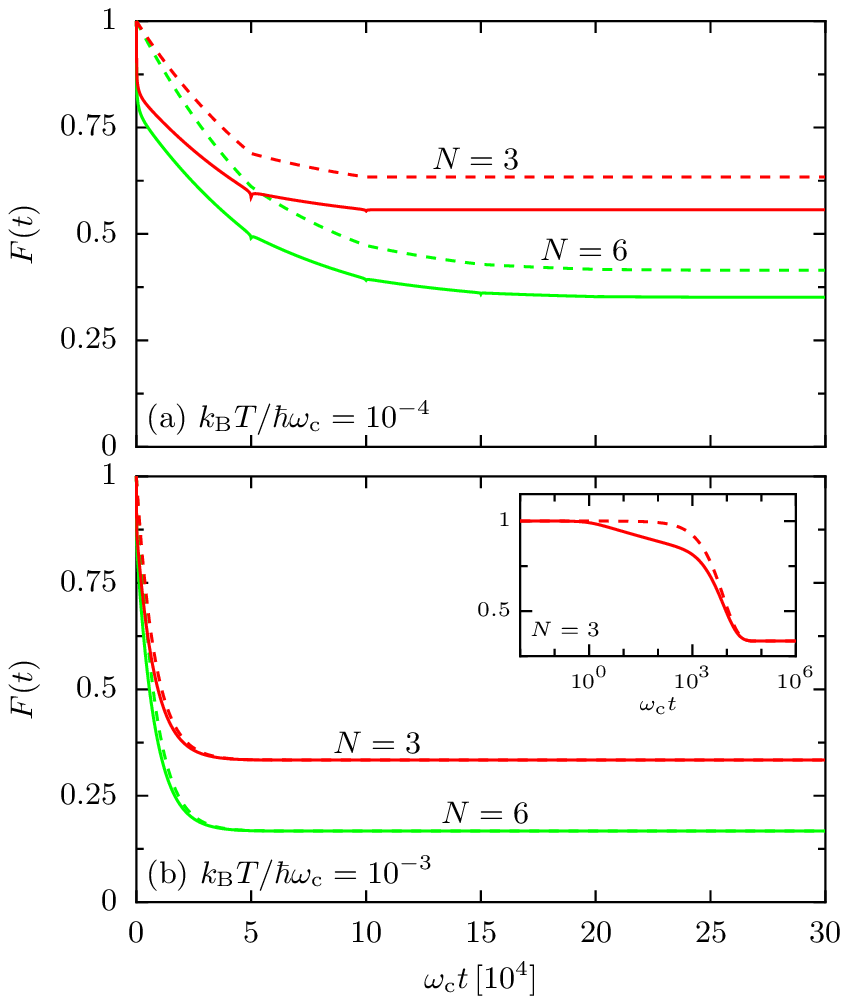}
\caption{\label{figFidelity}(Color online) Exact time evolution of the
fidelity $F(t)$ (solid lines) and the result obtained from the causal
master equation (dashed lines) for $N = 3$ and $N=6$ qubits,
respectively.  The temperatures are $k_\text{B}T =10^{-4}/\hbar
\omega_\text{c}$ (a) and $k_\text{B}T=10^{-3}/\hbar
\omega_\text{c}$ (b). As in Fig.~\ref{figCExactPhaseNoise},
the transit time between nearest-neighbor qubits is
$t_{12} = 5\times 10^4/\omega_\text{c}$ and the qubit-field coupling
strength $\alpha=0.005$.  The inset in panel (b) shows the data for
$N=3$ on a logarithmic time axis.}
\end{figure}

For the robust state~\eqref{eqnRobustStateNQubits}, the fidelity 
becomes
\begin{equation}\label{Frobust}
{F}(t) = \frac{1}{N}\sum_{j,j^\prime=1}^{N} \tilde
\rho_{jj^\prime}(t),
\end{equation}
where the coherences $\tilde\rho_{jj'}$ are given in
Eq.~\eqref{eqnMatrixElementsRobustStateRenormalized}. The time evolution of the
fidelity for $N=3$ and for $N=6$ qubits is shown in Fig.~\ref{figFidelity}. For
low temperatures [panel (a)], we find that the fidelity decay is slowed down
whenever a transit time is reached, i.e.~at times $t=t_{jj'}$. This resembles
the behavior of the concurrence for two qubits shown in
Fig.~\ref{figCExactPhaseNoise}. For a larger number of qubits, the fidelity
saturates at a lower value.
%-------------
%-------------

In order to gain a more quantitative understanding of the fidelity
saturation, we focus on the final fidelity $F(\infty)$. In the
zero-temperature limit, which provides a lower bound
for the coherence loss, we find from
Eqs.~\eqref{eqnMatrixElementsRobustStateRenormalized} and \eqref{exactexplicit} the density matrix elements
$\tilde\rho_{jj^\prime}(\infty) =
[1+\omega_\text{c}^{2}t_{12}^{2}(j-j^\prime)^{2}]^{-4\alpha}/N$. 
Hence the fidelity~\eqref{Frobust} saturates to 
\begin{equation}\label{eqnFidelityinfty}
F(\infty)
=\frac{1}{N} \Biggl[ 1 + 2 \sum_{q=1}^{N-1}\frac{1-q/N}{(1 +
q^{2}\omega_\text{c}^{2}t_{12}^{2})^{4\alpha}}\Biggr].
\end{equation} 
Although we start out with a ``robust'' entangled state, this final
fidelity $F(\infty)$ can be as low as $1/N$, which marks the large-distance 
limit $t_{12}\to\infty$.  More generally, we find that for
zero temperature, the final fidelity decreases with increasing cutoff
frequency $\omega_\text{c}$, increasing spatial separation, and for a
larger qubit-bath coupling strength $\alpha$.

An intriguing aspect of the fidelity is its scaling
behavior as a function of the system size $N$. Will $F(\infty)$ decay to zero for larger arrays of qubits, or converge to a finite value?
For large $N$ we can neglect the term $1/N$ in Eq.~(\ref{eqnFidelityinfty}) and replace the sum over $q$ by an
integration over the continuous variable $x=q/N$. Then we obtain 
\begin{eqnarray}\label{fidelityintegral}
F(\infty)
& \simeq & 2 \int_{0}^{1}\mbox{d}x\,\frac{1-x}{(1 + N^{2}
  \omega_\text{c}^{2}t_{12}^{2}x^{2})^{4\alpha} } \nonumber \\
&=& 2\;{ }_{2}F_{1}\left(\frac{1}{2},4\alpha,\frac{3}{2},-N^{2}
    \omega_\text{c}^{2}t_{12}^{2}\right) 
\nonumber \\
&&+  \frac{1 - (1+N^{2}\omega_\text{c}^{2}t_{12}^{2})^{1-4\alpha} }
{N^{2}\omega_\text{c}^{2}t_{12}^{2}(1-4\alpha)}
,
\end{eqnarray}
where the evaluation of the integral yields Gauss' hypergeometric function
${ }_{2}F_{1}$. 
The expression~(\ref{fidelityintegral}) is valid for general coupling constant $\alpha$, but beyond weak coupling its value is rather small. 
We will now approximate Eq.~(\ref{fidelityintegral}) for $\alpha \lll 1$.
Furthermore, usually many cutoff wavelengths $2\pi c/\omega_{\rm c}$ will fit
between two neighboring qubits, so that $\omega_\text{c}t_{12} \gg 1$, as
we argued above. Since we already assumed $N \gg 1$ to arrive at the
integral~\eqref{fidelityintegral}, we are surely in the limit $N
\omega_\text{c}t_{12}\gg 1$. Then we can approximate the
hypergeometric function by its asymptotic expansion for large
fourth argument. We finally obtain 
\begin{equation}\label{eqnFinftyapp}
F(\infty)
\simeq \left[\frac{\Gamma(\frac{1}{2}-4\alpha)}{\Gamma(\frac{3}{2}-4\alpha)}-\frac{1}{1-4\alpha}\right]   (N \omega_{c}t_{12})^{-8\alpha}.
\end{equation} 
Hereby we found the important result that although the final fidelity
is smaller for larger systems, the scaling is only algebraic in $N$.
This stability under dephasing is a property of the initial $N$-qubit W state
(\ref{eqnRobustStateNQubits}).
Clearly, for nonvanising $t_{12}$ this state lives in a
decoherence-poor rather than in a decoherence-free subspace.
Equation~(\ref{eqnFinftyapp}) shows that at zero temperature, the final
fidelity is determined by two dimensionless numbers, the one number
being $\alpha$ and the other the ratio between the array length $N c
t_{12}$ and the cutoff wavelength $2\pi c/\omega_\text{c}$.  

\begin{figure}
\includegraphics{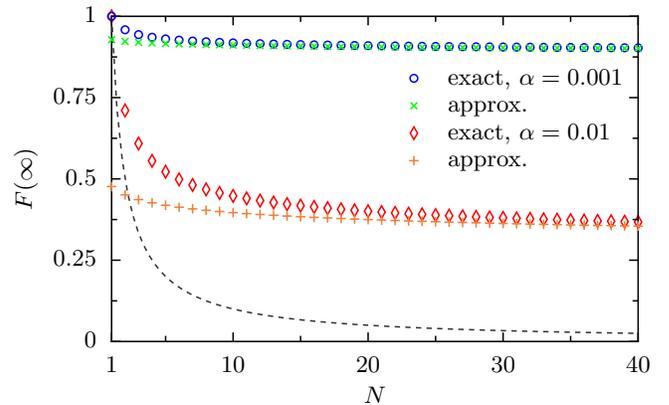}
\caption{\label{figFidelitySS}(Color online) Final fidelity \eqref{eqnFidelityinfty}
as a function of the number $N$ of qubits at zero temperature
for two coupling strengths $\alpha = 0.001$ (blue circles) and $\alpha = 0.01$
(red diamonds). The nearest-neighbor transit time is again $\omega_\text{c}
t_{12}= 5\times10^4$. The crosses (orange) and plus signs (green) mark the approximative result~\eqref{eqnFinftyapp},
respectively. The dashed line indicates the lower bound $1/N$ of the fidelity.}
\end{figure}

In Fig.~\ref{figFidelitySS}, we compare for two values of $\alpha$ the exact
expression~\eqref{eqnFidelityinfty} for the final fidelity as a function of $N$
with the weak-coupling approximate result~(\ref{eqnFinftyapp}). Clearly, for
state-of-the-art well-isolated qubits with typically $\alpha =0.001$, the
agreement is excellent already for $N\apprge 5$, while for $\alpha =0.01$
convergence is reached for $N\apprge 10$. The figure clearly shows that in the
weak coupling limit $\alpha\lll 1$, the final fidelity $F(\infty)$ in
(\ref{eqnFinftyapp}) is almost independent of the length of the array. 
For $\alpha < 0.005$ the factor in square brackets in Eq.~(\ref{eqnFinftyapp})
is less than $1.05$, so that the large-$N$ expression for the final fidelity could be
further simplified as $F(\infty)\simeq (N \omega_{c}t_{12})^{-8\alpha}$.
%-----------
%-----------

The above estimates were derived in the limit of strictly zero temperature, so
that the question arises up to which temperature they still represent a
reasonably good approximation.  A closer inspection of the exact
result~\eqref{exactexplicit} reveals that this is certainly the case if the
condition
\begin{equation}\label{lowT_Nqubits}
\frac{\hbar c}{k_{\rm B}T} \gg N c t_{12} 
\end{equation}
holds, i.e.~if the thermal coherence length of the bath is much larger than the
length of the array. Assuming $T=\unit[10]{mK}$, the thermal coherence length is
$\unit[2.3]{\mu m}$. This value corresponds to an array length of $24$ qubits
with a nearest-neighbor distance $x_{12}= c t_{12} = \unit[100]{nm}$. 
The final fidelity that we obtain for $T=0$ in Eq.~\eqref{eqnFinftyapp} can 
therefore be considered as an upper bound for what could be realized in state-of-the-art
quantum information processing experiments on arrays of qubits.

%------------------------------------------------------------------------------
\section{Causal master equation}\label{secMasterEqn}

In Sec.~\ref{secExact}, we found that the analytical
solution for the qubit dynamics can involve rather complex expressions and,
thus, an intuitive picture of the observed behavior can be hard to find,
even though the exact solution is known.  Thus for a more qualitative
understanding, one can benefit from an approximative treatment in the
spirit of a Bloch-Redfield master equation approach. Moreover, such an
approach enables a symmetry analysis of the dissipative time evolution.
This can provide additional insight in cases in which tracing out the
bath degrees of freedom reveals symmetries that are obeyed by the
dissipative central system, but not by the system-bath Hamiltonian.

Bloch-Redfield equations are based on a perturbative treatment
of the qubit-environment coupling, followed by neglecting memory
effects in the kernel of the resulting quantum master equation.
Thereby one entirely ignores the dependence of the dynamical
equations on the qubits' history and, thus, on the initial
preparation.  The resulting master equation then assumes the structure
$d\tilde\rho_{\mathbf{m,n}}/dt = \sum_{\mathbf{m',n'}}
R_{\mathbf{mnm'n'}} \tilde\rho_{\mathbf{m',n'}}$.
Such equations, however, fail to reproduce the stepwise
decay of the concurrence observed in Sec.~\ref{secExact}.
For the concurrence $C(t)$ of the robust Bell state
\eqref{eqnRobustState2Qubits}, they even yield $dC/dt = 0$ for all
distances $x_{12}$, in clear contrast to the exact solution. In other
words, strictly Markovian master equations can predict spurious
decoherence-free subspaces. In this section, we shall derive a master
equation that does not suffer from such a shortcoming. 

\subsection{Markov approximation and beyond}
Taking the above considerations as a motivation, we now derive a
generalization of the Bloch-Redfield master equation that is able to
capture the retardation effects stemming from a finite sound velocity.
In doing so we pursue closely the standard approach that leads to a
Markovian equation of motion. In this way, it can be identified where it fails
and how to improve it accordingly.

Starting with the Liouville-von Neumann equation
\eqref{eqnLiouvillevonNeumann}, we employ a projection-operator
formalism to formally eliminate the bath degrees of freedom. In
second-order perturbation in the qubit-bath coupling, we obtain for
the reduced qubit density matrix the master equation
\begin{equation} \label{eqnWeakCouplingMasterEqn}
\frac{d}{d t} \tilde\rho(t)
= - \sum_{\nu,\nu^\prime = 1}^N \int_0^{t-t_0}\diff\tau\,
  \kappa_{\nu\nu^\prime}(t,\tau) \,\tilde\rho(t)\,,
\end{equation}
where $t\ge t_0$ and the super-operator
\begin{equation}
\begin{split} \label{eqnMEIntegralKernel}
\kappa_{\nu\nu^\prime}(t,\tau)[\,\cdot\,]
&= \mathcal{S}_{\nu\nu^\prime}(\tau)
\bigl[\widetilde X_\nu(t), \bigl[ \widetilde X_{\nu^\prime}(t-\tau) ,\,\cdot\,
\bigr]\bigr]\\
& \quad + \im\mathcal{A}_{\nu\nu^\prime}(\tau) \bigl[ \widetilde X_\nu(t)\bigl\{ \widetilde
X_{\nu^\prime}(t-\tau) ,\, \cdot\, \bigr\}\bigr]\,,
\end{split}
\end{equation}
with the anti-commutator $\{\cdot,\cdot\}$.
For a sketch of the derivation, see Appendix~\ref{secWeakCouplingME}.
Although this master equation is in a
time-convolutionless form, it still is non-Markovian owing to the
explicit dependence on the initial time $t_0$. 
The integral kernel \eqref{eqnMEIntegralKernel} features the real part
$\mathcal{S}_{\nu\nu^\prime}(\tau)$ and the imaginary part
$\mathcal{A}_{\nu\nu^\prime}(\tau)$ of the bath correlation function
$\trace_\text{b}[\tilde\xi_\nu(\tau)\xi_{\nu^\prime}\rho^\text{eq}_\text{b}]$.
For our bath model they can be evaluated explicitly and read
\begin{align} \label{eqnCorrelationFunctionS}
\mathcal{S}_{\nu\nu^\prime}(\tau) &= \frac{1}{2} \left[
\mathcal{S}(\tau-t_{\nu\nu^\prime}) +  \mathcal{S}(\tau+t_{\nu\nu^\prime})\right]\,, \\ 
\label{eqnCorrelationFunctionA}
\mathcal{A}_{\nu\nu^\prime}(\tau) &= \frac{1}{2} \left[
\mathcal{A}(\tau-t_{\nu\nu^\prime}) +  \mathcal{A}(\tau+t_{\nu\nu^\prime})\right]\,,
\end{align}
where
\begin{align} 
\label{eqnStandardCorrelationFunctionS}
\mathcal{S}(\tau) &= \int_0^\infty\diff\omega\, J(\omega) \cos(\omega t)\coth\left(\frac{\hbar
\omega }{2 k_\text{B} T}\right)\,,\\ \label{eqnStandardCorrelationFunctionA}
\mathcal{A}(\tau) &= - \int_0^\infty \diff\omega\, J(\omega)\sin( \omega t )\,,
\end{align}
are the usual symmetric and anti-symmetric bath correlation
functions,\cite{Leggett1987a,vanKampen2001a,Breuer2002a} respectively.

We first consider the local terms, i.e.\ those with
$\nu=\nu'$, for which the time shift $t_{\nu\nu'}$ vanishes.
Then the correlation functions \eqref{eqnCorrelationFunctionS} 
and \eqref{eqnCorrelationFunctionA} reduce to Eqs.~\eqref{eqnStandardCorrelationFunctionS} and
\eqref{eqnStandardCorrelationFunctionA}, respectively, and
one can introduce a Markov approximation in the usual way: If the correlation functions
$\mathcal{S}(\tau)$ and $\mathcal{A}(\tau)$ contribute to the integral
in Eq.~\eqref{eqnWeakCouplingMasterEqn} essentially in a small time
interval of size $\tau_\text{b}$ around $\tau =
0$, then for $t-t_0 \gg \tau_\text{b}$, we can extend the
$\tau$-integration to infinity, i.e.\ we set
\begin{equation} \label{eqnMarkovLocal}
\int_0^{t-t_0}\diff\tau\, \kappa_{\nu\nu}(t,\tau)  \approx 
\int_0^{\infty}\diff\tau\, \kappa_{\nu\nu}(t,\tau)\,.
\end{equation}
This expression implies a coarse-graining in time so that the resulting master
equation is valid only for time steps not smaller than the bath correlation time
$\tau_\text{b}$. In general, the bath correlation time depends on the
properties of the spectral density $J(\omega)$ and the temperature of the
bath. If the temperature is not too low and the spectral density is fairly
smooth and decays sufficiently fast for $\omega \to 0$ and $\omega \to
\infty$, as is the case here, then the correlation time is only weakly
temperature dependent and reads $\tau_\text{b} \approx
1/\omega_\text{c}$.\cite{Cheng2005a}

In the above treatment of the local terms $\nu = \nu'$,
we have followed the route towards a Markovian equation of motion.
For the nonlocal terms, however, the arguments of the
last paragraph are no longer valid. For $\nu \neq \nu^\prime$ the
correlation functions \eqref{eqnCorrelationFunctionS} and
\eqref{eqnCorrelationFunctionA} are not peaked
at $\tau=0$, but at $\tau=t_{\nu\nu'}$ which for a realistic qubit
separation typically exceeds the bath correlation time
$\tau_\text{b}$.  Then we have to distinguish the cases
$t-t_0<\tau_\text{b}$ and $t-t_0>\tau_\text{b}$.  In the former case,
the peak of the correlation functions lies outside the integration
interval so that the integral is small and, consequently, will be
neglected. In the latter case, by contrast, the peak fully contributes
so that the integral can again be extended to infinity.  In summary, this
means
\begin{equation}\label{eqnMarkovNonlocal}
\begin{split}
& \int_0^{t-t_0}\diff\tau \kappa_{\nu\nu^\prime}(t,\tau) \\
&\qquad \approx \Theta(t-t_0-t_{\nu\nu^\prime})\int_0^{\infty} \diff \tau\,
\kappa_{\nu\nu^\prime}(t,\tau),
\end{split}
\end{equation}
where $\Theta(t)$ is the Heaviside step function.  For $\nu=\nu'$,
this expression coincides with Eq.~\eqref{eqnMarkovLocal}.

Inserting the approximation \eqref{eqnMarkovNonlocal} into the
weak-coupling master equation \eqref{eqnWeakCouplingMasterEqn} and
setting again the initial time $t_0 = 0$, we find
the {causal master equation} (CME)
\begin{equation}
\frac{d}{dt} \tilde\rho(t) = R(t)\tilde\rho(t)
\end{equation}
with the time-dependent superoperator
\begin{equation}
\label{R_CME}
\begin{split}
R(t)[\cdot]
= -\sum_{\nu,\nu^\prime = 1}^N & \Theta (t - t_{\nu\nu^\prime})
  \int_0^\infty \diff\tau
\\ \times
\Big( &\mathcal{S}_{\nu\nu^\prime}(\tau) \bigl[\widetilde X_\nu(t),
       \bigl[ \widetilde X_{\nu^\prime}(t-\tau) ,\,\cdot\, \bigr]\bigr]
\\
     + &i\mathcal{A}_{\nu\nu^\prime}(\tau) \bigl[ \widetilde X_\nu(t),\bigl\{
       \widetilde X_{\nu^\prime}(t-\tau) ,\, \cdot\, \bigr\}\bigr]
\Big) .
\end{split}
\end{equation}
The step functions ensure causality which requires that the cross
terms can only be active after the propagation time between the
respective qubits has passed. In the limit of vanishing separation, 
the causal master equation reduces to a standard Bloch-Redfield equation.
In the following, we will demonstrate that the causal master equation
reproduces the results of Sec.~\ref{secExact} rather well, while a
standard Bloch-Redfield approach clearly fails.

\subsection{Master equation for pure dephasing}\label{secCMEPD}
Let us now apply the causal master equation to the problem defined in
Sec.~\ref{secModel} and test the results against the exact solutions
presented in Sec.~\ref{secExact}. Since $X_\nu = \sigma_{\nu z}$
commutes with the free Hamiltonian $H_0$, the interaction-picture coupling
operators stay time-independent, $\widetilde X_\nu(t) = \sigma_{\nu z}$.
Then the time integration in the causal Bloch-Redfield tensor \eqref{R_CME}
involves only the bath correlation functions and we obtain for the density
matrix element $\tilde\rho_{\mathbf{m},\mathbf{n}}$ the equations of motion
\begin{equation}
\label{CME}
\frac{d}{dt}\tilde\rho_{\mathbf{m},\mathbf{n}}
= \Bigl[- \Lambda^\text{CME}_{\mathbf{m,n}}(t) + \im
\phi^\text{CME}_\mathbf{m,n}(t)\Bigr]
\, \tilde \rho_{\mathbf{m},\mathbf{n}}
\end{equation}
with the damping
\begin{align} \label{eqnCMEDamping}
\begin{split}
\Lambda^\text{CME}_{\mathbf{m,n}}(t) &= \frac{\alpha\pi k_\text{B}T}{\hbar}
  \sum_{\nu,\nu'=1}^N \Theta(t-t_{\nu\nu'})\\
  & \quad \times \bigl[ (-1)^{m_\nu + m_{\nu^\prime}} + (-1)^{n_\nu +
  n_{\nu^\prime}} \\
  & \quad\qquad - (-1)^{m_\nu+n_{\nu^\prime}} - (-1)^{m_{\nu^\prime} + n_\nu}
  \bigr],
\end{split}
\intertext{and the phase shift} \label{eqnCMEPhaseShift}
\begin{split}
\phi^\text{CME}_\mathbf{m,n}(t) &=  \sum_{\nu,\nu'=1}^N \Theta(t-t_{\nu\nu'})
\frac{\alpha\omega_\text{c}}{1+\omega_\text{c}^2 t_{\nu\nu'}^2} \\
 &\qquad \times \bigl[ (-1)^{m_\nu + m_{\nu^\prime}} - (-1)^{n_\nu + n_{\nu^\prime}}\bigr] .
 \end{split}
\end{align}
This master equation is non-Markovian due to the appearance of the
step functions, which change the effective damping and phase shift
whenever a transit time $t_{\nu\nu^\prime}$ is reached. These stepwise
time-dependent frequency shifts and decay rates are characteristic
features of the causal master equation. As we will see, because of
these steps the CME follows more closely the smooth time-dependent
variations of shifts and decay rates of the exact dynamics  than the
standard master-equation formalism manages to do with its static
shifts and decay rates.

As in Sec.~\ref{secExact}, we first consider as an initial
state the robust Bell state \eqref{eqnRobustState2Qubits}.
From the master equation \eqref{CME}, we find that the concurrence
$C_\text{CME} = 2|\tilde\rho_{01,10}|$ obeys
\begin{equation} \label{eqnMECRobust}
\frac{d}{d t} C_\mathrm{CME} = - \frac{8\alpha\pi
k_\text{B}T}{\hbar} \left[ 1 - \Theta (t-t_{12}) \right] \,
C_\mathrm{CME}\,.
\end{equation}
This differential equation is readily integrated to provide the solution
\begin{equation} \label{eqnSolutionMECRobust}
C_\mathrm{CME}(t) =
\begin{cases}
\e^{-8\alpha\pi k_\text{B}T t/\hbar}\,,& \quad 0 \leq t < t_{12} \\
\e^{-8\alpha\pi k_\text{B}T t_{12}/\hbar} = \text{const.}\,,&\quad t
\ge t_{12}\,,
\end{cases}
\end{equation}
i.e.\ the concurrence decays exponentially until the transition time
is reached and thereafter remains constant. This clear separation of
two dynamical regimes facilitates an intuitive interpretation: For
times $t<t_{12}$, the qubits have not ``seen'' each other, so that we are in a
regime of single-qubit decoherence. Indeed, during this first time interval 
the causal master equation~(\ref{CME}) coincides  with a standard Bloch-Redfield
approach in which the qubits are coupled to independent heat baths.
Consequently, the relative phase between the qubits is randomized and 
the concurrence decays. However, for $t>t_{12}$, both qubits experience
correlated quantum noise and undergo collective decoherence.
Thus, the concurrence decay comes to a standstill and a decoherence-poor 
subspace can emerge.

This time evolution is compared to the exact solutions in
Fig.~\ref{figCExactPhaseNoise}. We find that generally the causal master
equation describes the slow decay of the concurrence and its saturation very
well. At very short times, however, the causal master equation does not capture
the initial slip of the concurrence. The reason for this is, that the dynamics
on timescales that are comparable to the bath correlation time cannot be
resolved in a coarse-grained time approximation underlying the causal master
equation. The same generally holds true for Markovian quantum master equations,
in particular for the standard Bloch-Redfield equation. At larger times, the
benefits of the causal master equation become obvious: Since the Bloch-Redfield
treatment is recovered by setting the transit time $t_{12}=0$,
Eq.~\eqref{eqnMECRobust} reveals that $\diff C_\text{BR}/\diff t = 0$ for all
times, i.e.~the concurrence remains at its initial value. This spurious
robustness of the concurrence is, however, in clear contrast to the exact
result.
For pure phase noise, the populations of the system
eigenstates are conserved and, consequently, the final state is generally
unrelated to thermal equilibrium. Therefore, the question whether the
master equation can describe the grand canonical ensemble of the
system coupled to the bath is in the present context ill-posed.

\begin{figure}
\includegraphics{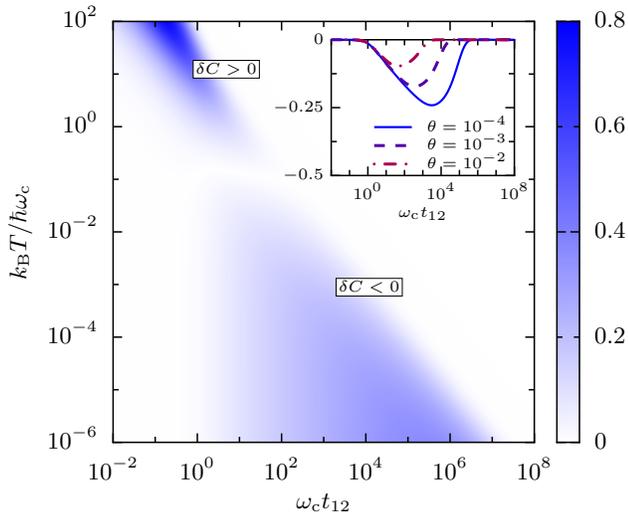}
\caption{\label{figCssExVsMePhaseNoise}(Color online) Difference
between the exact result and the causal master equation result
\eqref{eqnSolutionMECRobust} for the final
value of the concurrence, $\delta C = C(\infty) -
C_\text{CME}(\infty)$, for the robust entangled 2-qubit state
\eqref{eqnRobustState2Qubits}.
The coupling strength is $\alpha = 0.005$.  The inset depicts $\delta C$
for the fixed temperatures $\theta = k_\text{B}T / \hbar\omega_\text{c} =
10^{-4}$ (solid), $\theta = 10^{-3}$ (dashed) and $\theta = 10^{-2}$
(dash-dotted), respectively.}
\end{figure}

For a quantitative investigation of the quality of the causal
master equation, we compare the final values of the
concurrence, $\lim_{t\to\infty}C(t)$.  Figure~\ref{figCssExVsMePhaseNoise}
depicts the difference of the exact solution and the causal
master equation result, $\delta C = C(\infty)- C_\text{CME}(\infty)$,
as a function of transit time and temperature. For realistic temperatures
$T \lesssim 10^{-1} \hbar \omega_\text{c} / k_\text{B}$,
Fig.~\ref{figCssExVsMePhaseNoise} shows that the final value obtained from the
causal master equation exceeds the exact result, so that
$C_\text{CME}(\infty)$ provides an upper bound for the concurrence.
In particular, for $100 \,\hbar / k_\text{B} T \apprle t_{12} \apprle
\omega_\text{c}^{-1}$, the agreement is almost perfect. Note that for very high temperatures
$T \gtrsim 10^{-1}\hbar\omega_\text{c} / k_\text{B}$, the difference
$\delta C$ assumes also positive values.
%-------------
%-------------

Let us finally apply the master equation \eqref{CME} also to the case of a
linear $N$-qubit arrangement with equal nearest-neighbor spacings $x_{12}$ as
discussed in Sec.~\ref{secNQubits}.  We again consider the initial preparation
in the robust state \eqref{eqnRobustStateNQubits} using the shorthand notation
\eqref{eqnMatrixElementOneQubitExited}. To calculate the fidelity $F$ defined in
Eq.~\eqref{Frobust}, we need to compute the values of the density matrix
elements $\tilde\rho_{jj'}$. Unlike for the two-qubit concurrence, phase shifts
now also play a role. The time-dependent phase shift $\phi^\text{CME}_{jj'}(t)$
[see Eq.~\eqref{eqnCMEPhaseShift}] can be written as the difference of two
terms, the one only depending on $j$ and the other only on $j'$. Both terms
describe stepwise time-dependent frequency shifts of the corresponding qubits.
Interestingly, after the longest transit time $t>t_{1N}$, these shifts in the
CME become static and $\phi^\text{CME}_{jj'}(t)$ agrees with the exact
result~\eqref{eqnFrequencyShiftStatic}, i.e.\ $\phi_{jj'}^\text{CME}(t) =
\delta\Omega_j - \delta\Omega_{j^\prime}$. For an unambiguous comparison of
fidelities in the exact and in the causal master-equation formalism, it is an
important result that we can work in the same interaction picture with the same
renormalized frequencies $\Omega_j \to \Omega_j + \delta \Omega_j$. It remains
to be discussed what is the effect of the stepwise frequency shifts in the CME
for times $t<t_{1N}$. It is the non-diagonal terms $\nu\neq\nu'$ in
Eq.~\eqref{eqnCMEPhaseShift} that make up the difference between the frequency
shifts at time $t=0$ and the exact static renormalization at times $t> t_{1N}$.
However, this difference is very small due to large factors
$\omega_\text{c}^{2}t_{\nu\nu'}^{2} \gg 1$ in the denominator of
Eq.~\eqref{eqnCMEPhaseShift}, and can safely be neglected in the following.

From the causal master equation \eqref{CME} we then obtain
\begin{equation}\label{eqnMEMatrixElementNRobust}
\frac{d}{d t}\tilde\rho_{jj^\prime}(t) = - \frac{8\alpha\pi
k_\text{B}T}{\hbar} [ 1 -
\Theta(t-t_{jj^\prime})] \tilde\rho_{jj^\prime}(t)\,.
\end{equation}
Interestingly enough, in the present case only two types of terms of the
master equation~\eqref{CME} contribute: the local terms with $\nu = \nu^\prime$ and those with
$|\nu-\nu^\prime| = |j-j^\prime|$. As a consequence, the decay rate of $\rho_{jj'}(t)$
changes only at the transit time $t_{jj^\prime}$.

In order to evaluate the fidelity~\eqref{Frobust}, we integrate
Eq.~\eqref{eqnMEMatrixElementNRobust} and sum over all density matrix
elements $\tilde\rho_{jj^\prime}(t)$, so that we obtain
\begin{equation} \label{eqnMENQubitFidelity}
\begin{split}
F_\text{CME}(t)
= \frac{1}{N^2} \sum_{j,j^\prime = 1}^N \big[&
  \Theta(t_{jj^\prime} -t) \e^{-8\alpha\pi k_\text{B}T t/\hbar}\\
+ & \Theta(t-t_{jj^\prime}) \e^{-8\alpha\pi k_\text{B}T
  t_{jj^\prime}/\hbar}\big]\,.
\end{split}
\end{equation}
For vanishing qubit separation, the master equation predicts
$F_\text{CME}(t)=1$, i.e.\ a decoherence-free behavior.  For $x_{12}
> 0$, however, all coherences $\tilde\rho_{jj^\prime}$ initially decay
and so does the fidelity.  When the smallest transit time
$t_{12}$ is reached, a perfect
correlation between nearest neighbors is built up and the
coherences $\tilde\rho_{j,j+1}$ saturate. Since these $N-1$ coherences are 
no longer time-dependent, the fidelity decay is reduced accordingly. 
This process continues until ultimately all transition times have passed, 
i.e.\ until $t = t_{1N}$, and the fidelity decay comes to a standstill.

In Fig.~\ref{figFidelity}, we compare this behavior to the exact solution.  For
very low temperatures [panel (a)], we observe that the master equation
reproduces the reduction of the fidelity decay whenever a transit time is
reached. The relative difference between the exact result and the causal master
equation is of the order of 10\% as in the case of two qubits. If the
temperature becomes larger, so that the nearest neighbor separation exceeds the
thermal coherence length [see Fig.~\ref{figCssExactPhaseNoise} and the
discussion after Eq.~\eqref{eqnSaturatedConcurrence}], then the fidelity decay
is determined by thermal noise. In that case, $F$ already saturates to a rather
small value before the first transit time is reached. The inset of
Fig.~\ref{figFidelity}b shows that at very short times when the decay sets in,
i.e.~when the CME is not yet different from a standard master equation for
independently dephasing qubits, that then the CME result can deviate
significantly from the exact result. Again we wish to stress that this
deviation is due to the coarse-grained time approximation that is inherent in a
standard master equation approach as well. However, in contrast to the latter, 
the CME and the exact results agree very well at later times.

%----------------------------------------------------------------
\section{Discussion and conclusions}\label{Sec:Conclusions}

Decoherence of an array of qubits can be a rather complex process that proceeds
in several qualitative different stages, even in the case of pure phase noise
which we investigated here: If the qubits are coupled to the bosonic field of
the substrate vibrations, the dynamics during a first, very short period is
essentially noiseless. In a second stage, the bosonic vacuum fluctuations are
most relevant, while finally, thermal noise dominates. The spatial separation of
the qubits brings in further time scales, namely the transit times of sound
waves between the qubits.

Here we considered robust entangled $N$-qubit W states, which do not decohere
for vanishing qubit separations. For finite separations we find instead that the
qubits start to dephase. However, the dephasing slows down whenever the elapsed
time reaches a transit time, until it eventually comes to a standstill. The
final $N$-qubit quantum coherence increases with decreasing qubit-qubit
separation, qubit-bath coupling strength, cutoff frequency, and temperature. By
contrast, single qubits in the same one-dimensional environment would loose all
quantum coherence for all finite values of these bath parameters.
Note that, the two-qubit W state is identical to the robust-entangled Bell
state. The saturation of entanglement that we find does not occur for
fragile two-qubit states.\cite{Doll2006a,Solenov2007a} 

Cooperative effects can be advantageous or detrimental, depending on the
specific protocol that one has in mind. For example, one may fight decoherence
by creating  decoherence-free subspaces. To that end, one could bring the qubits
close together and use the W states for quantum information processing because
their entanglement is robust. Nevertheless, the qubits must be sufficiently well
separated, either to enable their individual manipulation or because of their
finite extensions. We found that this requirement prevents the realization of
decoherence-free subspaces. Cooperative effects are still advantageous, since
decoherence-poor subspaces are built up instead. Good results require the length
of the array of qubits to be smaller than the thermal coherence length $\hbar
c/k_\text{B}T$.

Alternatively, one may wish to implement active quantum error-correction
schemes, where logical qubits are redundantly encoded into several physical
qubits. Here, by contrast, the cooperative effects are detrimental, since
standard error-correction schemes\cite{Nielsen2000a,Klesse2005a} and recent
generalizations to non-Markovian baths\cite{Terhal2005a,Aliferis2006a} will only
work perfectly if the physical qubits couple to spatially uncorrelated baths.
Our discussion demonstrates that neighboring physical qubits should then be
separated by more than the thermal coherence length $\hbar c/k_\text{B}T$. By
reducing the temperature, single-qubit dephasing is suppressed, but the
assumption of uncorrelated baths becomes worse. This suggests that there may be
an optimal working temperature for quantum error correction, given a geometry of
physical qubits.

Thus our calculations show how well decoherence-free subspaces or quantum error
correction-protocols could be realized with linear arrays of qubits. The
aforementioned conflicting requirements for both strategies seem to rule out the
implementation of both strategies in one experiment.

One should keep in mind that bit-flip noise, which was not considered here, will
reduce coherence further, although typically on a longer time scale. Including
bit-flip noise usually renders the qubits-environment model no longer exactly
solvable. Therefore one has to resort to approximation schemes like, e.g, a
master-equation approach. As we have shown, the common Bloch-Redfield master
equation cannot account for the intrinsically non-Markovian effects stemming
from the spatial separations. For the present model, the Bloch-Redfield approach
predicts spurious decoherence-free subspaces, which in fact are at best
decoherence-poor. Very recently, it was found\cite{Karasik2007a} that for
bit-flip noise due to a homogeneous isotropic Markovian and {\em
three}-dimensional reservoir, there is no multi-particle decoherence-free
subspace outside the Dicke limit, i.e.\ whenever the qubits are not co-located.
The situation may be different for one-dimensional structures for which we find
that at least for pure dephasing, decoherence-free subspaces can become
imperfect if the qubits are separated.

In order to capture delocalization effects with a master equation, we have
derived a modified Bloch-Redfield approach that ensures causality for the
qubit-qubit interaction mediated by the substrate. It proved to be reliable for
parameters for which the standard master equation for a single qubit is
reliable. This is the case for sufficiently high temperatures or small enough
coupling strengths such that initial-slip effects are small. A characteristic
feature of the proposed causal master equation is that it selects the
Bloch-Redfield kernel depending on the time elapsed since the preparation. This
means that the time evolution is governed by a time-dependent Liouville operator
which renders the dynamics non-Markovian.  Besides being a proper tool for
studying retardation effects in models that do not possess an exact solution,
the causal master equation describes the time evolution in an intuitive and
concise manner. Thereby, it enables decoherence studies with algebraic methods
which possibly will provide suggestions for coherence stabilization.

In this sense, our studies of the interplay between pure dephasing and a spatial
qubit separation can only be a first step towards a deeper understanding of such
phenomena.  In particular the inclusion of other quantum noise sources, which
are also present in real experiments, will complement the picture drawn above.

% ----------- acknowledgements -----------------------------------------------
\begin{acknowledgments}
This work was supported by Deutsche Forschungsgemeinschaft through
SFB~484 and SFB~631. PH and SK acknowledge funding by the DFG
excellence cluster ``Nanosystems Initiative Munich''.
%
%This work was supported by Deutsche Forschungsgemeinschaft through
%SFB~484, SFB~631, and the excellence cluster ``Nanosystems Initiative
%Munich''.
\end{acknowledgments}
%------------------------------------------------------------------------------
\appendix

\section{Exact reduced dynamics}\label{secDerivationExactSolution}

In this appendix, we outline the derivation of the exact
solution~\eqref{eqnExactSolution} for the reduced qubit dynamics for
the case of pure dephasing.\cite{Palma1996a, Duan1998a, Reina2002a,
Doll2006a}.  In the usual interaction picture with respect to the
uncoupled qubits and the bath, the coupling Hamiltonian
\eqref{eqnCoupling} reads
\begin{equation}
\widetilde H_\mathrm{qb}(t) = \widetilde V(t) + \widetilde V^\dagger(t)\,,
\end{equation}
with the interaction
\begin{equation}
\widetilde V(t) = \hbar \sum_\nu \sigma_{\nu z} \sum_k g_{k} b_k
\e^{\im(kx_\alpha - \omega_k t)}\,.
\end{equation}
Our aim is to calculate the time evolution of the reduced density matrix
$\tilde\rho(t) = \trace_\mathrm{b} \tilde R(t) = \trace_\mathrm{b} [ U(t) R(0)
U^\dagger(t) ]$ generated by the propagator
\begin{equation} \label{eqnPropagator}
U(t) = \mathcal{T} \exp\left( \frac{1}{\im \hbar} \int_0^t \!\diff s\,
\widetilde H_\mathrm{qb}(s)\right ),
\end{equation}
where $\mathcal{T}$ denotes the time-ordering operator. In a first step,
we find
$[\widetilde V(t), \widetilde V^\dagger(t^\prime)] = f(t-t^\prime)$,
where
\begin{equation}
f(t) = \hbar^2 \sum_k\sum_{\nu\nu^\prime} g^2_{k}  \sigma_{\nu z}\sigma_{\nu^\prime z} \e^{\im (kx_{\nu\nu^\prime} - \omega_k t)}\,.
\end{equation}
Since $[\widetilde V(t), \widetilde V(t^\prime)]\,{=}\,[\widetilde V^\dagger(t),
\widetilde V^\dagger(t^\prime)]\,{=}\,0$ and $[f(t), \widetilde
V(t^\prime)]$ $=[f(t), \widetilde V^\dagger(t^\prime)]\,{=}\,0$ for all times $t$ and $t^\prime$,
we can use the Baker-Campbell-Hausdorff formula\cite{Gardiner1991a} to
express the time-ordered exponential \eqref{eqnPropagator} as
\begin{equation} \label{eqnPropagator2}
\begin{split}
U(t)
= \exp\bigg\{ & \frac{1}{\im \hbar} \int_0^t\!\diff s\, \widetilde
  H_\mathrm{qb}(s)
  -\frac{1}{\hbar^2}\int_0^t\!\diff s\int_0^t \!\diff s^\prime\,
\\ \times
& f(s-s^\prime) \left[\theta(s-s^\prime) - \theta(s^\prime - s) \right] 
\bigg\}\,.
\end{split}
\end{equation}
The first term in the exponent can be written as
\begin{equation} \label{eqnPropagatorFirstExponential}
\exp\left\{\frac{1}{\im \hbar} \int_0^t \widetilde H_\mathrm{qb}(t)\,\diff
s\right\} = \prod_k D_k \Bigl(\sum_\nu \sigma_{\nu z} y_{\nu k}\Bigr)\,,
\end{equation}
where we defined $y_{\nu k} = g_k \e^{-\im k x_\nu} (1-\e^{\im
\omega_k t})/\omega_k$ and the displacement operators $D_k(X) = \exp\{X
b_k^\dagger - X^\dagger b_k\}$.
The second term in the exponent is a c-number and provides the
time-dependent phase factor $\exp[\im\phi(t)]$ with
\begin{equation}
\phi(t) = \sum_k \sum_{\nu,\nu^\prime=1}^N g^2_{k} \frac{\omega_k t - \sin(\omega_k t)}{\omega_k^2} 
\sigma_{\nu z} \sigma_{\nu^\prime z} \e^{\im k
x_{\nu\nu^\prime}}  \,.
\end{equation}
So far we have found for the propagator the expression
\begin{equation}\label{eqnPropagator3}
U(t) = \prod_k D_k \Bigl(\sum_\nu \sigma_{\nu z} y_{\nu k}\Bigr)
e^{\im \phi(t)} .
\end{equation}
For the factorizing initial condition $R(0) =
\rho(0)\rho^\mathrm{eq}_{\mathrm{b}}$, the matrix elements of the
reduced density operator become
\begin{equation}\label{eqnDensityMatrixElement}
\begin{split}
\tilde\rho_{\mathbf{m},\mathbf{n}}(t) &= \trace_\mathrm{b} \langle \mathbf{m} | U(t)
\rho(0)\rho^\mathrm{eq}_{\mathrm{b}} U^\dagger(t) | \mathbf{n} \rangle\\
&= \rho_{\mathbf{m},\mathbf{n}}(0) \trace_\mathrm{b} \left\{ \rho^\mathrm{eq}_{\mathrm{b}} \langle \mathbf{n} |
U^\dagger(t) | \mathbf{n} \rangle  \langle \mathbf{m} | U(t) | \mathbf{m} \rangle
 \right\}\,.
\end{split}
\end{equation}
In the second equality of Eq.~\eqref{eqnDensityMatrixElement} we have
used the cyclic property of the trace and the fact that the
computational basis elements $|\mathbf{n}\rangle$ are eigenstates
of the propagator. Inserting the propagator \eqref{eqnPropagator3}
into \eqref{eqnDensityMatrixElement} and assuming an isotropic coupling
$g_{-k} = g_k$, we find for density matrix element
$\tilde\rho_{\mathbf{m},\mathbf{n}}(t)$ the phase
\begin{align}
\phi_{\mathbf{m},\mathbf{n}}(t) &= \langle \mathbf{m} | \phi(t) | \mathbf{m}
\rangle - \langle \mathbf{n} | \phi(t) | \mathbf{n} \rangle\\
\begin{split}
&=\sum_k\sum_{\nu, \nu^\prime}^N g^2_{k} \frac{\omega_k t - \sin(\omega_k
t)}{\omega_k^2} \\
&\qquad \times\left[ (-1)^{m_\nu + m_{\nu^\prime}} - (-1)^{n_\nu +
n_{\nu^\prime}}\right]  \cos( k x_{\nu\nu^\prime})\,,
\end{split}
\end{align}
which in the continuum limit becomes the phase~\eqref{eqnPhaseShift}.
For the calculation of the remaining contributions in
\eqref{eqnDensityMatrixElement}, we employ the relations
$D^\dagger_k(X) = D_k(-X)$ and
\begin{equation}\label{eqnProductOfDisplacements}
D_k(X)D_k(Y) = \e^{(XY^\dagger - X^\dagger Y)/2}  D_k(X+Y) ,
\end{equation}
which hold for any commuting operators $X$ and $Y$.
Then we obtain
\begin{align} \label{eqnDensityMatrixElement2a}
\tilde\rho_{\mathbf{m},\mathbf{n}} & (t) \nonumber \\
={} &
 \rho_{\mathbf{m},\mathbf{n}}(0)\,\e^{\im \phi_{\mathbf{m},\mathbf{n}}(t)}
\\ \nonumber & \times
 \trace_\mathrm{b} \Big\{ \rho^\mathrm{eq}_{\mathrm{b}} \langle \mathbf{n} |
 \prod_k D_k^\dagger \big(\sum_\nu \sigma_{\nu z} y_{\nu k}\big)
 |\mathbf{n}\rangle
\\ \nonumber & \times
 \langle\mathbf{m}|
 \prod_k D_k \bigl(\sum_\nu \sigma_{\nu z} y_{\nu k}\bigr) |\mathbf{m}\rangle
 \Big\}
\\ \label{eqnDensityMatrixElement2}
={}&
 \rho_{\mathbf{m},\mathbf{n}}(0)\,\e^{\im [\phi_{\mathbf{m},\mathbf{n}}(t) +
 \eta_{\mathbf{m},\mathbf{n}}(t)] }\\ &\times
 \prod_k \trace_\mathrm{b} \Big\{ \rho^\mathrm{eq}_{\mathrm{b},k} D_k
 \big( \sum_\nu \left[ (-1)^{m_\nu} - (-1)^{n_\nu} \right] y_{\nu k} \big)
 \Big\} . \nonumber
\end{align}
An additional phase $\eta_{\mathbf{m},\mathbf{n}}(t)$ stems from the
commutator of the displacement operators $D_k$ in
\eqref{eqnDensityMatrixElement2a} [see Eq.~\eqref{eqnProductOfDisplacements}]
and reads
\begin{equation}
\begin{split}
\eta_{\mathbf{m},\mathbf{n}}(t) &= 2\sum_k\sum_{\nu,\nu^\prime=1}^N g_k^2
\frac{1-\cos(\omega_k t)}{\omega_k^2}\\
&\qquad \times (-1)^{n_{\nu^\prime} + m_{\nu}} \sin(k x_{\nu\nu^\prime})
\end{split}
\end{equation}
and vanishes for isotropic coupling $g_{-k} = g_k$ which we assume herein.

Finally, we have to evaluate the trace in Eq.~\eqref{eqnDensityMatrixElement2}.
This can be accomplished conveniently in the basis of the coherent
states $|\beta_k\rangle$, in which the equilibrium bath density
operator \eqref{rhobath} reads
\begin{equation} \label{eqnCoherentStateRepresentation}
\rho_{\mathrm{b},k}^\mathrm{eq} = \frac{1}{\pi \langle n_k \rangle} \int
\!\diff^2\beta_k\, \e^{-|\beta_k|^2/\langle n_k\rangle} |\beta_k\rangle \langle \beta_k |\,.
\end{equation}
The integration is over the whole complex plane and $\langle n_k \rangle =
[\exp(\hbar \omega_k/k_\mathrm{B} T) -1]^{-1}$ is the Bose
distribution function.  $\beta_k$ are the complex eigenvalues of the
annihilation operator $b_k$, i.e.~$b_k
|\beta_k\rangle = \beta_k | \beta_k\rangle$. After inserting
expression \eqref{eqnCoherentStateRepresentation} into
Eq.~\eqref{eqnDensityMatrixElement2}, we integrate for each mode $k$
over the complex plane. We finally end up with
\begin{equation}
\tilde\rho_{\mathbf{m},\mathbf{n}}(t) =
\rho_{\mathbf{m},\mathbf{n}}(0)\,\e^{\im \phi_{\mathbf{m},\mathbf{n}}(t) -
\Lambda_{\mathbf{m},\mathbf{n}}(t)}\,,
\end{equation}
where
\begin{equation}
\begin{split}
\Lambda_{\mathbf{m},\mathbf{n}}(t) &= \sum_k g_k^2 \frac{1 - \cos(\omega_k
t)}{\omega_k^2} \coth\left( \frac{\hbar\omega_k}{2k_\mathrm{B} T}\right) \\
&\qquad \times \left | \sum_{\nu=1}^N \left[ (-1)^{m_\nu} - (-1)^{n_\nu} \right]
\e^{\im k x_\nu} \right|^2\,.
\end{split}
\end{equation}
Note that $\Lambda_{\mathbf{m},\mathbf{n}}(t)$ is real-valued and thus
accounts for the damping of the matrix element. In the continuum
limit, it assumes the form~\eqref{eqnDamping}.

%----------------------------------------------------------
\section{Weak-coupling master equation}\label{secWeakCouplingME}

For notational convenience, we write the Liouville-von Neumann
equation \eqref{eqnLiouvillevonNeumann} with the super-operator
$\tilde{\mathcal{L}}(t)[\,\cdot\,] = \lambda [\tilde
H_\mathrm{qb}(t), \,\cdot\,] / \im \hbar $, so that it reads
$\partial_t \tilde R(t) = \tilde{\mathcal{L}}(t) \tilde R(t)$,
where $\lambda$ will serve as an expansion parameter.
The formal solution can be written as $\tilde R(t) =
\mathcal{U}(t,t_0) \tilde R(t_0)$, where
\begin{equation}\label{eqnLiouvilleSuperOperator}
\begin{split}
\mathcal U(t,t_0) &= \Biggl[1 + \int_{t_0}^t \tilde{\mathcal{L}}(t_1) \,\diff t_1 \\
&\qquad + \int_{t_0}^{t} \int_{t_0}^{t_1} \tilde{\mathcal{L}}(t_1) 
\tilde{\mathcal{L}}(t_2)\,\diff t_1 \diff t_2 \, + \dots \Biggr]\,.
\end{split}
\end{equation}
Moreover, we introduce the projection operator
\begin{equation} \label{eqnProjectionOperator}
\langle \,\cdot\,\rangle  = \trace_\mathrm{b}
\{\,\cdot\,\}\otimes\rho_\mathrm{b}^\mathrm{eq}
\end{equation}
which projects an operator of the full Hilbert space (of the qubits
and the bath) to a direct product of a system operator and
a time-independent reference state $\rho_\mathrm{b}^\mathrm{eq}$ of
the bath.  Here, we are interested in the relevant part $\langle\tilde
R(t)\rangle$ of the density matrix, which describes the
reduced dynamics. The projection operator
\eqref{eqnProjectionOperator} is constructed such that the factorizing
initial condition~\eqref{eqnInitialCondition} provides the identity
$\langle \tilde R(t_0)\rangle = \tilde R(t_0)$. This allows one to
write the formal solution of the dissipative system dynamics in the form
\begin{equation} \label{eqnLiouvilleReduced}
\langle \tilde R(t)\rangle = \langle \mathcal{U}(t,t_0) \rangle
\tilde R(t_0)\,.
\end{equation}

From Eq.~\eqref{eqnLiouvilleReduced}, one can derive a formally exact
quantum master equation for $\langle \tilde R(t) \rangle$ in two
ways:
The first possibility is the Nakajima-Zwanzig projector
technique\cite{Nakajima1958a, Zwanzig1960a} which leads to an
integro-differential equation for $\langle \tilde R \rangle$ which is
nonlocal in time. The second possibility, on which we will focus
here, is termed ``time convolutionless projection operator
technique''\cite{Kubo1962a, vanKampen1974a, Breuer2002a} 
and leads to a equation of motion of the form
\begin{equation} \label{eqnTCLMasterEquation}
\frac{d}{d t} \langle\tilde R(t)\rangle = \mathcal{K}(t)
\langle\tilde R(t)\rangle\,,
\end{equation}
which is local in time, but is governed by an explicitly
time-dependent super-operator $\mathcal{K}(t)$.  Although this
equation possesses an apparently simple form, it generally cannot be
solved exactly and, thus, one has to resort to a perturbative
treatment. In doing so, we expand the generator $\mathcal{K}(t)$ in
powers of the qubit-bath coupling parameter $\lambda$,
i.e.~$\mathcal{K}(t) = \sum_n \lambda^n \mathcal{K}_n(t)$. Then a
formal integration of Eq.~\eqref{eqnTCLMasterEquation} results in
\begin{equation} \label{eqnTCLMasterEquationIntegrated}
\begin{split}
\langle \tilde R(t) \rangle = \Biggr[ 1 &+ \lambda \int_{t_0}^t
\mathcal{K}_1(t_1) \,\diff t_1 +\lambda^2 \Biggl(  \int_{t_0}^t\mathcal{K}_2(t_1)\,\diff t_1\\
& + \int_{t_0}^t \int_{t_0}^{t_1} \mathcal{K}_1(t_1)
\mathcal{K}_1(t_2) \,\diff t_2 \diff t_1 \Biggr) + \ldots \Biggr]
\tilde R(t_0) .
\end{split}
\end{equation}
Comparing the powers of $\lambda$ of Eqs.~\eqref{eqnLiouvilleReduced}
and \eqref{eqnTCLMasterEquationIntegrated}, we find the relations
\begin{align}
\mathcal{K}_1(t) &= \langle \tilde{\mathcal L}(t) \rangle\,,\\
\mathcal{K}_2(t) &= \int_{t_0}^t \langle \tilde{\mathcal L}(t) \tilde{\mathcal
L}(t_1) \rangle - \langle \tilde{\mathcal{L}}(t)\rangle\langle \tilde{\mathcal
L}(t_1)\rangle\,\diff t_1\,,
\end{align}
so that the master equation reads
\begin{equation} \label{eqnTCLMasterEquation2ndOrder}
\frac{d}{d t} \langle\tilde R(t)\rangle
= \langle \tilde{\mathcal L}(t) \rangle \,\langle\tilde R(t)\rangle
 +\int_{t_0}^t \diff t_1 \langle \tilde{\mathcal L}(t) \tilde{\mathcal L}(t_1)
  \rangle \langle\tilde R(t)\rangle\,.
\end{equation}
By inserting the definitions of the projection operator
\eqref{eqnProjectionOperator} and the Liouvillian
$\tilde{\mathcal{L}}$, we obtain Eq.~\eqref{eqnWeakCouplingMasterEqn}.
Note that the coupling~\eqref{eqnCoupling} results in $\langle
\tilde{\mathcal L}(t) \rangle = 0$.

\end{document}